\documentclass[a4paper]{article}
\usepackage{arxiv}

\usepackage[utf8]{inputenc} 
\usepackage[T1]{fontenc}    
\usepackage{hyperref}       
\hypersetup{
  pdfinfo={
    Title={Past as a Guide: Leveraging Retrospective Learning
for Python Code Completion},
    Author={Seungyoun Shin, Seunggyu Chang, Sungjoon Choi },
    Subject={If you want to put something here, do so},
    Keywords={Add some keywords if you feel so inclined}
  }
}
\usepackage{url}            
\usepackage{booktabs}       
\usepackage{amsfonts}       
\usepackage{nicefrac}       
\usepackage{microtype}      
\usepackage{xcolor}         
\usepackage{graphicx}
\usepackage{tikz}
\usepackage{comment}
\usepackage{amsmath,amssymb} 
\usepackage{color}
\usepackage{mathtools}
\usepackage{listings}
\usepackage{algorithm}
\usepackage{algpseudocode}
\usepackage{fancyvrb}
\usepackage{mdframed}
\usepackage{tcolorbox}
\usepackage{lipsum}
\usepackage{tabularx}
\usepackage{array}
\usepackage{colortbl}
\usepackage{caption}
\usepackage{tabularx}
\usepackage{pdfpages}
\usepackage{listings}
\usepackage{xcolor}

\newcommand{\cmark}{{\color{green}\checkmark}}
\newcommand{\xmark}{{\color{red}\sffamily X}}

\definecolor{codegreen}{rgb}{0,0.6,0}
\definecolor{codeblue}{rgb}{0.25,0.5,0.75}
\definecolor{codegray}{rgb}{0.5,0.5,0.5}
\definecolor{backcolour}{rgb}{0.95,0.95,0.95}

\definecolor{mylightblue}{rgb}{0.93,0.95,1.0}
\definecolor{mygray}{rgb}{0.86,0.86,0.86}
\newcolumntype{A}{>{\centering\arraybackslash}m{9em}}

\lstdefinestyle{mystyle}{
  backgroundcolor=\color{backcolour},   
  commentstyle=\color{codegreen},
  keywordstyle=\color{magenta},
  numberstyle=\tiny\color{codegray},
  stringstyle=\color{codeblue},
  breakatwhitespace=false,         
  breaklines=true,                 
  captionpos=b,                    
  keepspaces=true,                 
  numbers=left,                    
  numbersep=5pt,                  
  showspaces=false,                
  showstringspaces=false,
  showtabs=false,                  
  tabsize=2
}
\tcbset{
  modernrect/.style={
    enhanced,
    drop shadow={blur shadow},
    colback=white,
    colframe=blue!75!black,
    fonttitle=\bfseries,
    leftright skip=1.5cm,
    beforeafter skip=1cm,
    title=Modern Rectangle Box,
    attach boxed title to top left={yshift=-0.25cm,xshift=1cm},
    boxed title style={
      sharp corners, 
      colback=blue!75!black, 
      colframe=blue!75!black
    },
    before upper={\hspace*{1cm}},after upper={\hspace*{1cm}},
  }
}

\title{
Past as a Guide: Leveraging Retrospective Learning for Python Code Completion
}

\author{
  Seunggyoon Shin\textsuperscript{1,2}, Seunggyu Chang\textsuperscript{2}, Sungjoon Choi\textsuperscript{1} \\
  \textsuperscript{1} Korea University \\
  \textsuperscript{2} NAVER Cloud \\
}

\begin{document}

\maketitle

\begin{abstract}
This work presents Past as a Guide (PaG), a simple approach for Large Language Models (LLMs) to improve the coding capabilities by integrating the past history with interactive and iterative code refinements.
To be specific, inspired by human cognitive processes, the proposed method enables LLMs to utilize previous programming and debugging experiences to enhance the Python code completion tasks. 
The framework facilitates LLMs to iteratively refine the Python code based on previous execution and debugging results and optimize learning and reasoning capabilities. 
The proposed methodology achieved a 92\% pass@1 on HumanEval, demonstrating the potential to advance the field by leveraging retrospection from past experiences and interactive and iterative refinement processes without external correctness indicators.
\end{abstract}

\section{Introduction}
\label{intro}

Large Language Models (LLMs) play a pivotal role in advancing artificial intelligence research, exhibiting versatility across various applications including coding interfaces and search mechanisms. This versatility is emphasized by seminal works such as ReAct~\cite{yao2022react}, SayCan~\cite{ahn2022can}, Toolformer~\cite{schick2023toolformer}, among others. Concurrently, leveraging code is becoming a substantial medium for improving task performance, as illustrated by PAL~\cite{gao2023pal} and Program of Thoughts~\cite{chen2022program}. Our study, "Past as a Guide(PaG)," aligns with these advancements, aiming to optimize LLM performance through the integration of interactive and iterative code refinement from past retrospective guidance.

This mirrors a fundamental trait of human cognition: the intrinsic ability to harness past experiences. By adeptly employing short-term and long-term memory encoding and retrieval, humans address complex reasoning tasks, especially in areas like programming. Consider, for instance, the task of finding the last 100 digits of the n-th Fibonacci number. Often, when confronted with such challenges, humans draw from their reservoir of past experiences. They recall similar tasks, such as computing the n-th Fibonacci number, and intuitively gauge the similarities and differences between the old and new challenges. This process demonstrates a synergy: humans use long-term memory retrieval to construct retrospections from prior experiences and concurrently engage in short-term iterative refinements. 

We conduct empirical evaluations to assess the performance of LLMs in harnessing retrospections, leading to a significant achievement of 92\% pass@1 at HumanEval. Instead of relying on direct external feedback on correctness, we utilize the MBPP dataset ~\cite{austin2021program} to construct a long-term memory for the LLMs. When faced with a problem from HumanEval \cite{chen2021evaluating}, the model retrieves and leverages similar problems from its MBPP experiences, effectively boosting its performance. This method, combining interactive and iterative code refinements informed by past experiences, signifies a pivotal advancement in LLM capabilities, indicating a promising trajectory for their future evolution.

\begin{figure}[h!]
\centering
\includegraphics[width=0.8\textwidth]{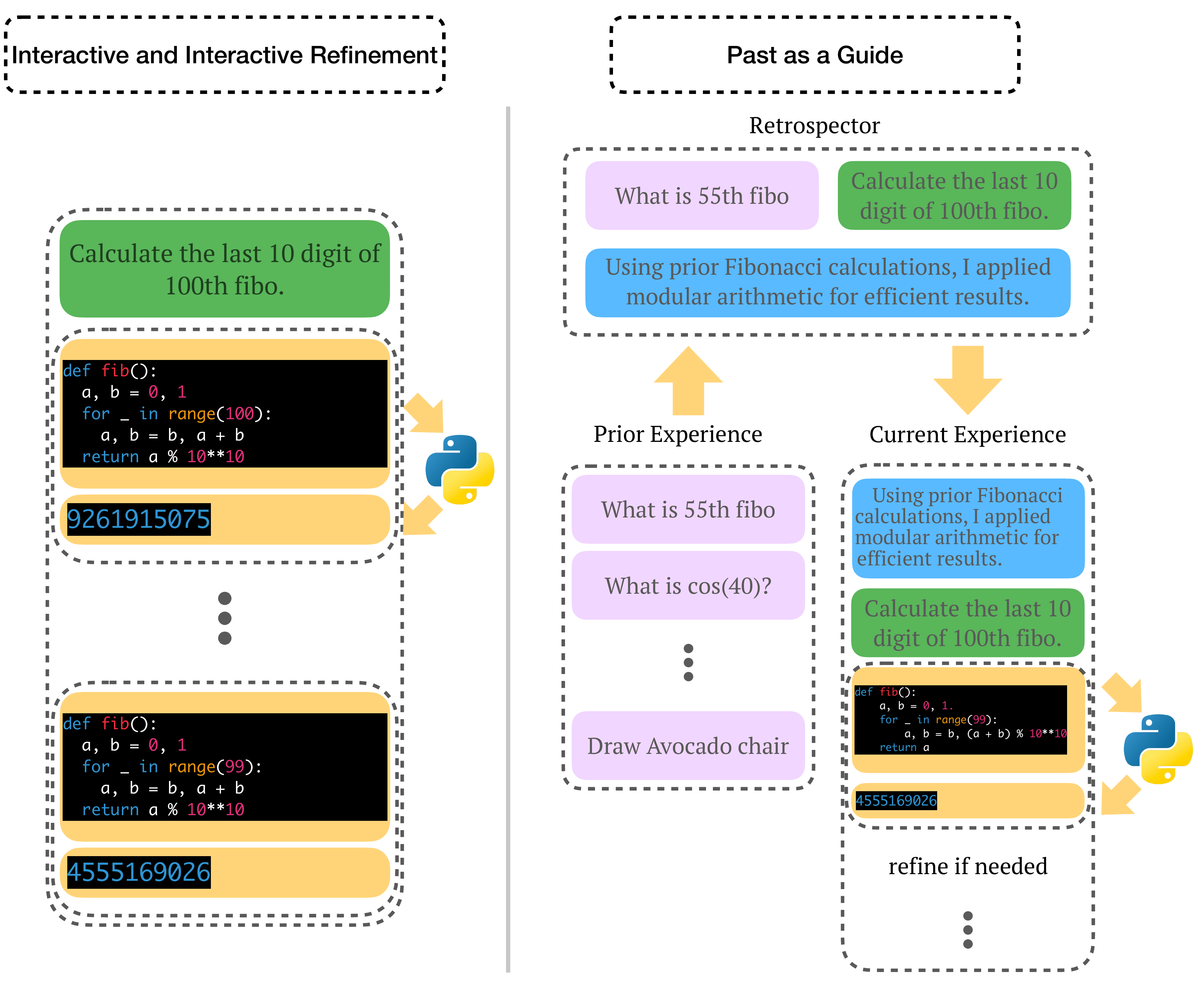}

\caption{Left: \textcolor{gray}{\textbf{Interactive and Iterative Code Refinement}} showcased through a process of repeated corrections. Right: Retrieval of the most similar question, aligned with the \textcolor{green}{\textbf{Current instruction}}, sourced from the \textcolor{pink}{\textbf{Past Experience}}. By merging these insights, we form \textcolor{blue}{\textbf{retrospections}}. The subsequent \textcolor{yellow}{\textbf{code generation}} follows an interactive and iterative Code refinement process, emphasizing code execution and debugging.}

\label{fig:fi1}
\end{figure}

\section{Related work}
\label{related_work}


\subsection{Code Generation LLMs}

There has been a surge of interest in generating code using LLMs. Notable advancements in this domain include the methodologies proposed by Roziere et al.~\cite{roziere2023code}, Li et al.~\cite{li2023starcoder}, Chen et al.~\cite{chen2021evaluating}, and another separate work by Li~\cite{li2023starcoder}. These research pieces underscore the capability of LLMs to craft precise and functional code structures. Additionally, innovative techniques have emerged that harness models like ChatGPT or those that deploy self-instruction paradigms~\cite{wang2022self}. Such models are first primed to generate paired datasets of instructions and code, which are subsequently used to fine-tune the LLMs~\cite{luo2023wizardcoder,honovich2022unnatural}. In a similar vein, Zhou et al.~\cite{zhou2022docprompting} leverage external knowledge through "DocPrompting" by integrating code documentation for code generation. While this approach underscores the value of external references, our approach uniquely focuses on leveraging the past experiences of LLMs to improve problem-solving in programming tasks.

\subsection{Code Execution LLMs}

An alternative but equally compelling line of work exploits code not merely as an output but as a vital operative tool for LLMs. For instance, the PAL ~\cite{gao2023pal} utilizes code execution results embedded within prompts to boost the model's proficiency in mathematical and numerical benchmarks. Moreover, the Binder approach~\cite{cheng2022binding} demonstrates that LLMs, when equipped with SQL or Python execution capabilities, manifest a superior comprehension of tabular datasets. Exploring the intersection of robotics and LLMs, the Code as Policies ~\cite{liang2023code} postulates that LLMs can produce code, serving as operational policies, to govern robotic actions fulfilling specific instructions. Furthermore, ViperGPT~\cite{suris2023vipergpt}, an innovative paradigm that synergizes vision foundation models with Python-executable code generated by GPT to address vision-oriented tasks.

\subsection{Self Improvement}

Various studies have highlighted innovative approaches to enhance the programming performance of Language Learning Models (LLMs). For instance, Haluptzok et al.~\cite{haluptzok2022language} demonstrated that crafting programming puzzles and validating solutions collaboratively can produce verified synthetic data. When LLMs are fine-tuned with this data, there's a marked improvement in their programming capabilities. On a different tangent, "Reflexion," as developed by Shinn et al.~\cite{shinn2023reflexion}, grants agents dynamic memory and the ability for self-reflection to improve reasoning abilities. Although Reflexion employs a form of external memory, its application is constrained to scenarios involving identical instruction sets. This methodology, while effective for repeated tasks, may not generalize well to varied problems that share conceptual similarities but differ in specifics, as is common in real-world applications. Consequently, in our comparative analysis, Reflexion is not categorized as using external memory in a broad sense, which we reserve for systems that leverage past insights dynamically across distinct but conceptually related problems. In the realm of mathematics, Zhou et al.~\cite{zhou2023solving} emphasize that prompts focused on verification, combined with voting techniques, can notably enhance GPT-4's performance in math-related benchmarks. Another intriguing approach is presented by Madaan et al.~\cite{madaan2023self}, wherein the model iteratively refines its outputs leveraging self-generated feedback. "Voyager" by Wang et al.~\cite{wang2023voyager} conceptualizes code as a skill. It recalls and incorporates skills acquired from prior tasks, thus sharpening its proficiency in abstractions. In contrast to the above methodologies that often rely on external rewards or feedback, our proposed PaG utilizes generated codes to rectify errors discerned from prior retrospections. Notably, our method constructs these retrospections solely based on its own execution results, without the need for any external input or feedback.

\begin{table}[htbp]
\centering
\caption[Comparison of LLM Enhancement Methodologies]{Comparison of LLM Enhancement Methodologies. *As discussed in ~\cite{huang2023large}, we do not classify receiving right or wrong feedback from an oracle evaluator as oracle feedback; instead, we utilize tools to obtain intrinsic feedback.}
\label{tab:comparison}
\begin{tabular}{@{}lccc@{}}
\toprule
& \textbf{Iterative Self Correction} & \textbf{External Memory*} & \textbf{No Oracle Feedback†} \\
\midrule
Voyager & \xmark & \cmark & \cmark \\
Self-refine ~\cite{madaan2023self} & \cmark & \xmark & \cmark \\
Reflexion ~\cite{shinn2023reflexion} & \cmark & \xmark* & \xmark \\
PaG (Ours) & \cmark & \cmark & \cmark \\
\bottomrule
\end{tabular}
\newline
\footnotesize{*External Memory is considered here as the ability to apply learned insights across varied but conceptually related problems. Reflexion's memory use is limited to identical instructions.}
\footnotesize{†In the context of oracle feedback, we refer to the use of explicit right/wrong evaluation from an oracle. Our approach utilizes intrinsic feedback mechanisms as delineated in ~\cite{huang2023large}.}
\end{table}

\section{Method}

\subsection{Interactive and Iterative Code Refinements}

\noindent \textbf{Overview:} Consider an instruction \( I \) presented in natural language. To achieve the desired outcome based on this instruction, we utilize a policy, denoted as \( \pi \). This policy, in the context of our work, is embodied by a Large Language Model (LLM) denoted as \texttt{LLM}. The policy \( \pi \) produces a verbal action \( A \) which, in essence, is textual content. When the content of \( A \) contains a code segment, it can be systematically parsed and executed. The outcome of this execution is subsequently appended to \( A \). Engaging in this process iteratively allows for the comprehensive completion of the task at hand.  
To explicate further, the action \( A \) is comprised of a set, such that:
\[ A = \{(\texttt{text}, \texttt{code}, \texttt{execution\_result})\} \times \text{\# of trials} \]
Here, each element of the set \( A \) represents a trial, constituting the text, the parsed code, and the result of executing that code, whose collective enumeration accomplishes the instructed task.

\noindent \textbf{Implementations:} For our experiment, we employ the GPT4 API. It has been observed that several popular LLM APIs, such as ChatGPT and GPT4, exhibit constraints when tasked with code generation. Primarily, many extant LLM models are trained with safety mechanisms to prevent potential harm, as discussed by Bai et al. (2022) ~\cite{bai2022constitutional}. Consequently, when requests are made for operations like fetching data from 'yfinance' or image manipulation tasks, these models might assert their inability to execute them. Contrarily, they are indeed capable of producing the correct code snippets necessary for task completion. To circumvent this limitation, we delineated the model's role and its execution capabilities within the \texttt{SYSTEM\_PROMPT} section of ChatGPT. A comprehensive version of this prompt can be found in Appendix ~\ref{sys_prompt}. Then, We utilized Python's regex functionality to extract and filter out code blocks. If a code segment is detected, it is executed using a Jupyter Notebook client. Subsequently, the resultant output from the code segment is assimilated into the GPT4 trajectories.
 
\subsection{Exploring stage}

\noindent \textbf{Memory Curation:} In the exploration stage, each run of an \texttt{LLM} produces a tuple, denoted by \( t \), and is mathematically represented as:
\[ t = \langle \texttt{USER\_INSTRUCTION}, \texttt{ASSISTANT\_ANSWER} \rangle \]

This tuple, \( t \), is then appended to a memory store, symbolized as \( \mathcal{M} \), in accordance with the following relation:
\[ \mathcal{M} \gets \mathcal{M} \cup \{t\} \]

Here, \( \mathcal{M} \) acts as a repository embodying the aggregate experiences or memories of the \texttt{LLM}. Each element within \( \mathcal{M} \) is a unique tuple comprising a user instruction and the corresponding assistant response. The incorporation of new tuples into \( \mathcal{M} \) signifies the \texttt{LLM}'s process of updating its knowledge base, learning, and adapting through each run. For the purpose of this paper, we leverage the initial subset of MBPP ~\cite{austin2021program}, consisting of the first 470 examples, to formulate the \( \mathcal{M} \)

\subsection{Past as a Guide}

PaG concentrates on utilizing experienced memories \( \mathcal{M} \) to proficiently formulate an action, \( A \), capable of fulfilling a provided instruction, \( I \). This method unfolds systematically through several interconnected steps. 

Initially, an instruction, \( I \), instigates a search within our structured memory store \( \mathcal{M} \). To facilitate this, we calculate embeddings for \( I \) and each instruction in \( \mathcal{M} \) using OpenAI's \texttt{text-embedding-ada-002} model and employ cosine similarity as the metric to quantify resemblance between the instruction embeddings, thereby extracting the most analogous instruction from \( \mathcal{M} \).

\begin{equation}
I_{\text{similar}} = \operatorname*{argmax}_{I' \in \mathcal{M}} \text{ Similarity}(I, I')
\end{equation}

When a similar instruction is identified, we use its corresponding action trajectories, \( A \), together with the original instruction, \( I \), to query the LLM. This query is designed to derive retrospections, denoted as \( R \). It emphasizes the extraction and application of knowledge from past experiences to the present context.

\begin{equation}
R = \texttt{LLM}\left(\text{“How can prior experience "} A_{\text{similar}} \text{" be applied to solve "} I \text{"?”}\right)
\end{equation}

After the acquisition of retrospections, \( R \), they are seamlessly integrated as a prefix to the initial instruction, \( I \), enriching the instruction base to generate the refined action, \( A \), targeting the accomplishment of the outlined task in the initial instruction.

\begin{equation}
A = \texttt{LLM}(R + I)
\end{equation}

PaG utilizes knowledge from previous interactions stored in \( \mathcal{M} \). This knowledge helps refine the model's responses to new instructions, placing retrospection of past experiences in the context of new situations. As a result, the model becomes more effective in completing tasks.

For a more detailed, practical illustration of our method refer to Appendix ~\ref{full_traj_example}.

\section{Code Completion Experiments}


\textbf{Task}: The task is derived from HumanEval \cite{chen2021evaluating}, where the objective is to complete a provided Python code skeleton. This task presented a natural language description along with an incomplete Python code snippet. The completed code is then verified for correctness by executing the associated test code, which either yields a pass or fail outcome based on the code’s accuracy and efficacy. The primary goal is to successfully fulfill the requirements delineated in the natural language description by augmenting the Python skeleton to create a functional and correct code piece.

\textbf{Results}: In our experiments, we assess the effectiveness of our proposed methods against baseline models. The following table 2. \ref{tab:performance_comparison} summarizes the performances of the different approaches:

\begin{table}[h]
\centering
\begin{tabular}{lc}
\toprule
Method & \% Pass@1\\
\midrule
GPT-4 & 67.00 \\
GPT-4 + † (max 6 tries) & 84.15 \\
GPT-4 + † (max 12 tries) & 90.85 \\
Reflexion ~\cite{shinn2023reflexion} & 91.00 \\
\textbf{GPT-4 + † (max 12 tries) + PaG} & \textbf{92.07} \\
\bottomrule
\end{tabular}
\caption{Performance comparison of different methods, with our method highlighted. The symbol † represents 'Interactive and Iterative Code Refinement'.}
\label{tab:performance_comparison}
\end{table}

PaG achieved state-of-the-art performance, a commendable feat realized without relying on external correctness feedback from the test code. Our approach uniquely combines retrospective guidance from past experiences with an interactive and iterative code refinements loop driven by code execution. This emphasizes the dual strengths of our method: harnessing past experiences for informed guidance and using interactive refinements for precision. Both factors collectively enhance the model's performance, showcasing the potential of integrating experienced knowledge with dynamic refinements without the need for external correctness indicators.

The actions generated by the Large Language Model, denoted as \( \mathcal{A} \), include numerous code blocks, each undergoing multiple trials. Every block within \( \mathcal{A} \) is subjected to a comprehensive examination against corresponding test cases. However, due to the possibility of discrepancies arising during intermediate executions of the solution function, an erroneous 'fail' flag may be prompted even if the final solution is accurate. To counteract this, we employ a refinement procedure, using regular expressions to meticulously filter and eliminate all print and assert statements within the generated code blocks. Any cases initially identified as failures undergo a re-assessment phase, wherein the accurate code block is extracted from the ensemble within \( \mathcal{A} \) and subjected to a secondary evaluation.

\section{Conclusion and Limitation}
\label{conclusion}

This work has developed a methodology that, akin to human learning, utilizes prior experiences and retrospection to enhance performance from the initiation phase through the reasoning stages, employing a multi-attempt code interpreter. Our proposed method, PaG, integrates past experiences with current instructions to craft retrospections. Further, it employs interactive and iterative code refinements through execution. By drawing on past lessons, PaG attained a 92\% performance on the HumanEval benchmark. However, the current scope of this methodology is largely confined to programming-centric tasks, similar to those in HumanEval, presenting a limitation in its applicability to more language-intensive tasks. As a prospective avenue for future research, the exploration of incorporating online learning to augment retrospections is promising, potentially allowing models to mirror the continual learning and adaptability inherent in human cognition and addressing a broader array of tasks with increased efficacy and precision.

%
%

\bibliographystyle{plain}
\bibliography{references}

\begin{thebibliography}{10}

\bibitem{ahn2022can}
Michael Ahn, Anthony Brohan, Noah Brown, Yevgen Chebotar, Omar Cortes, Byron David, Chelsea Finn, Chuyuan Fu, Keerthana Gopalakrishnan, Karol Hausman, et~al.
\newblock Do as i can, not as i say: Grounding language in robotic affordances.
\newblock {\em arXiv preprint arXiv:2204.01691}, 2022.

\bibitem{austin2021program}
Jacob Austin, Augustus Odena, Maxwell Nye, Maarten Bosma, Henryk Michalewski, David Dohan, Ellen Jiang, Carrie Cai, Michael Terry, Quoc Le, et~al.
\newblock Program synthesis with large language models.
\newblock {\em arXiv preprint arXiv:2108.07732}, 2021.

\bibitem{bai2022constitutional}
Yuntao Bai, Saurav Kadavath, Sandipan Kundu, Amanda Askell, Jackson Kernion, Andy Jones, Anna Chen, Anna Goldie, Azalia Mirhoseini, Cameron McKinnon, et~al.
\newblock Constitutional ai: Harmlessness from ai feedback.
\newblock {\em arXiv preprint arXiv:2212.08073}, 2022.

\bibitem{chen2021evaluating}
Mark Chen, Jerry Tworek, Heewoo Jun, Qiming Yuan, Henrique Ponde de~Oliveira Pinto, Jared Kaplan, Harri Edwards, Yuri Burda, Nicholas Joseph, Greg Brockman, et~al.
\newblock Evaluating large language models trained on code.
\newblock {\em arXiv preprint arXiv:2107.03374}, 2021.

\bibitem{chen2022program}
Wenhu Chen, Xueguang Ma, Xinyi Wang, and William~W Cohen.
\newblock Program of thoughts prompting: Disentangling computation from reasoning for numerical reasoning tasks.
\newblock {\em arXiv preprint arXiv:2211.12588}, 2022.

\bibitem{cheng2022binding}
Zhoujun Cheng, Tianbao Xie, Peng Shi, Chengzu Li, Rahul Nadkarni, Yushi Hu, Caiming Xiong, Dragomir Radev, Mari Ostendorf, Luke Zettlemoyer, et~al.
\newblock Binding language models in symbolic languages.
\newblock {\em arXiv preprint arXiv:2210.02875}, 2022.

\bibitem{gao2023pal}
Luyu Gao, Aman Madaan, Shuyan Zhou, Uri Alon, Pengfei Liu, Yiming Yang, Jamie Callan, and Graham Neubig.
\newblock Pal: Program-aided language models.
\newblock In {\em International Conference on Machine Learning}, pages 10764--10799. PMLR, 2023.

\bibitem{haluptzok2022language}
Patrick Haluptzok, Matthew Bowers, and Adam~Tauman Kalai.
\newblock Language models can teach themselves to program better.
\newblock {\em arXiv preprint arXiv:2207.14502}, 2022.

\bibitem{honovich2022unnatural}
Or~Honovich, Thomas Scialom, Omer Levy, and Timo Schick.
\newblock Unnatural instructions: Tuning language models with (almost) no human labor.
\newblock {\em arXiv preprint arXiv:2212.09689}, 2022.

\bibitem{huang2023large}
Jie Huang, Xinyun Chen, Swaroop Mishra, Huaixiu~Steven Zheng, Adams~Wei Yu, Xinying Song, and Denny Zhou.
\newblock Large language models cannot self-correct reasoning yet.
\newblock {\em arXiv preprint arXiv:2310.01798}, 2023.

\bibitem{li2023starcoder}
Raymond Li, Loubna~Ben Allal, Yangtian Zi, Niklas Muennighoff, Denis Kocetkov, Chenghao Mou, Marc Marone, Christopher Akiki, Jia Li, Jenny Chim, et~al.
\newblock Starcoder: may the source be with you!
\newblock {\em arXiv preprint arXiv:2305.06161}, 2023.

\bibitem{liang2023code}
Jacky Liang, Wenlong Huang, Fei Xia, Peng Xu, Karol Hausman, Brian Ichter, Pete Florence, and Andy Zeng.
\newblock Code as policies: Language model programs for embodied control.
\newblock In {\em 2023 IEEE International Conference on Robotics and Automation (ICRA)}, pages 9493--9500. IEEE, 2023.

\bibitem{luo2023wizardcoder}
Ziyang Luo, Can Xu, Pu~Zhao, Qingfeng Sun, Xiubo Geng, Wenxiang Hu, Chongyang Tao, Jing Ma, Qingwei Lin, and Daxin Jiang.
\newblock Wizardcoder: Empowering code large language models with evol-instruct.
\newblock {\em arXiv preprint arXiv:2306.08568}, 2023.

\bibitem{madaan2023self}
Aman Madaan, Niket Tandon, Prakhar Gupta, Skyler Hallinan, Luyu Gao, Sarah Wiegreffe, Uri Alon, Nouha Dziri, Shrimai Prabhumoye, Yiming Yang, et~al.
\newblock Self-refine: Iterative refinement with self-feedback.
\newblock {\em arXiv preprint arXiv:2303.17651}, 2023.

\bibitem{roziere2023code}
Baptiste Rozi{\`e}re, Jonas Gehring, Fabian Gloeckle, Sten Sootla, Itai Gat, Xiaoqing~Ellen Tan, Yossi Adi, Jingyu Liu, Tal Remez, J{\'e}r{\'e}my Rapin, et~al.
\newblock Code llama: Open foundation models for code.
\newblock {\em arXiv preprint arXiv:2308.12950}, 2023.

\bibitem{schick2023toolformer}
Timo Schick, Jane Dwivedi-Yu, Roberto Dess{\`\i}, Roberta Raileanu, Maria Lomeli, Luke Zettlemoyer, Nicola Cancedda, and Thomas Scialom.
\newblock Toolformer: Language models can teach themselves to use tools.
\newblock {\em arXiv preprint arXiv:2302.04761}, 2023.

\bibitem{shinn2023reflexion}
Noah Shinn, Federico Cassano, Beck Labash, Ashwin Gopinath, Karthik Narasimhan, and Shunyu Yao.
\newblock Reflexion: Language agents with verbal reinforcement learning.
\newblock {\em arXiv preprint arXiv:2303.11366}, 2023.

\bibitem{suris2023vipergpt}
D{\'\i}dac Sur{\'\i}s, Sachit Menon, and Carl Vondrick.
\newblock Vipergpt: Visual inference via python execution for reasoning.
\newblock {\em arXiv preprint arXiv:2303.08128}, 2023.

\bibitem{wang2023voyager}
Guanzhi Wang, Yuqi Xie, Yunfan Jiang, Ajay Mandlekar, Chaowei Xiao, Yuke Zhu, Linxi Fan, and Anima Anandkumar.
\newblock Voyager: An open-ended embodied agent with large language models.
\newblock {\em arXiv preprint arXiv:2305.16291}, 2023.

\bibitem{wang2022self}
Yizhong Wang, Yeganeh Kordi, Swaroop Mishra, Alisa Liu, Noah~A Smith, Daniel Khashabi, and Hannaneh Hajishirzi.
\newblock Self-instruct: Aligning language model with self generated instructions.
\newblock {\em arXiv preprint arXiv:2212.10560}, 2022.

\bibitem{yao2022react}
Shunyu Yao, Jeffrey Zhao, Dian Yu, Nan Du, Izhak Shafran, Karthik Narasimhan, and Yuan Cao.
\newblock React: Synergizing reasoning and acting in language models.
\newblock {\em arXiv preprint arXiv:2210.03629}, 2022.

\bibitem{zhou2023solving}
Aojun Zhou, Ke~Wang, Zimu Lu, Weikang Shi, Sichun Luo, Zipeng Qin, Shaoqing Lu, Anya Jia, Linqi Song, Mingjie Zhan, et~al.
\newblock Solving challenging math word problems using gpt-4 code interpreter with code-based self-verification.
\newblock {\em arXiv preprint arXiv:2308.07921}, 2023.

\bibitem{zhou2022docprompting}
Shuyan Zhou, Uri Alon, Frank~F Xu, Zhengbao Jiang, and Graham Neubig.
\newblock Docprompting: Generating code by retrieving the docs.
\newblock In {\em The Eleventh International Conference on Learning Representations}, 2022.

\end{thebibliography}


\appendix

\section{Code Interpreter System Prompt}
\label{sys_prompt}

\begin{tcolorbox}[title=\texttt{SYSTEM\_PROMPT} for Code Interpreter]

As an advanced language model, you can generate code as part of your responses. 
To make the code more noticeable and easier to read, please encapsulate it within triple backticks.

For instance, if you're providing Python code, wrap it as follows:

\begin{lstlisting}[language=Python]
print('hellow world')
\end{lstlisting}

Wrapped code block will automatically be executed and appended to the prompt.

\begin{lstlisting}[language=Python]
hellow world
\end{lstlisting}

After presenting the results from the code, you will provide a useful explanation or interpretation of the output to further aid your understanding. Additionally, when generating plots or figures, You’ll save them to a specified path, like ./tmp/plot.png so
that they can be viewed. After saving the plot, I’ll use the following markdown syntax to display the image at the end of the
response:

\begin{lstlisting}
![plot]('./tmp/plot.png')
\end{lstlisting}

You are using a Jupyter Notebook currently.
This approach allows me to visually present data and findings.

\end{tcolorbox}

\section{Retrospections Example Trajectory}
\label{full_traj_example}

\begin{tcolorbox}[title=\texttt{Retrospections}]
\textbf{From Previous Similar Questions:}\\
The \texttt{largest\_number} function sorts a list of digits in descending order and joins them to form the largest possible number. However, this is not directly applicable to the \texttt{rolling\_max} problem.
\end{tcolorbox}

\begin{tcolorbox}
[title=\texttt{Application to the Question}]
\textbf{Application to the Question:}\\
The \texttt{rolling\_max} function requires maintaining a rolling maximum value as we traverse through the list. This is different from the \texttt{largest\_number} function which sorts the entire list first. For \texttt{rolling\_max}, we need to keep track of the maximum value found so far and update it as we encounter larger numbers in the list. The output list should contain the maximum value found at each point in the input list. The \texttt{largest\_number} function does not provide a useful trajectory for solving this problem.
\end{tcolorbox}

\begin{tcolorbox}[title=\texttt{Instruction}]
Write a Python script to solve the following problem:

\begin{lstlisting}[language=Python]
from typing import List, Tuple

def rolling_max(numbers: List[int]) -> List[int]:
	# From a given list of integers, generate a list
	# of rolling maximum element found until given moment
	# in the sequence.
	>>> rolling_max([1, 2, 3, 2, 3, 4, 2])
	[1, 2, 3, 3, 3, 4, 4]
\end{lstlisting}

Ensure the solution is verified by printing the expected output.
\end{tcolorbox}

\begin{tcolorbox}[title=\texttt{Assistant}]
Here is the Python script that solves the problem:

\begin{lstlisting}[language=Python]
from typing import List

def rolling_max(numbers: List[int]) -> List[int]:
    max_so_far = numbers[0]
    result = []
    for num in numbers:
        if num > max_so_far:
            max_so_far = num
        result.append(max_so_far)
    return result

print(rolling_max([1, 2, 3, 2, 3, 4, 2]))
\end{lstlisting}

\begin{lstlisting}[language=Python]
RESULT
[1, 2, 3, 3, 3, 4, 4]
\end{lstlisting}

Additional Test Cases for Verification:

\begin{lstlisting}[language=Python]
print(rolling_max([1]))  # Single element
print(rolling_max([5, 5, 5, 5]))  # All elements are the same
print(rolling_max([9, 8, 7, 6, 5, 4, 3, 2, 1]))  # Descending order
print(rolling_max([]))  # Empty list
\end{lstlisting}

\begin{lstlisting}[language=Python]
RESULT
[1]
[5, 5, 5, 5]
[9, 9, 9, 9, 9, 9, 9, 9, 9]
Error: IndexError: list index out of range
\end{lstlisting}

Error Resolution and Final Implementation:

The function fails when the input list is empty, as the code tries to access the first element without prior existence verification. A conditional return for empty lists has been added to address this.

\begin{lstlisting}[language=Python]
from typing import List

def rolling_max(numbers: List[int]) -> List[int]:
    if not numbers:
        return []
    max_so_far = numbers[0]
    result = []
    for num in numbers:
        if num > max_so_far:
            max_so_far = num
        result.append(max_so_far)
    return result

# Re-verification with the adjusted implementation
print(rolling_max([]))  # Empty list
\end{lstlisting}

\begin{lstlisting}[language=Python]
RESULT
[]
\end{lstlisting}

The corrected implementation now successfully handles all edge cases, producing accurate results consistently.
\end{tcolorbox}

\end{document}